\documentclass[review]{elsarticle}

\usepackage{lineno,hyperref}
\modulolinenumbers[5]

\usepackage{graphicx}
\usepackage{dcolumn}
\usepackage{bm}
\usepackage{multirow}
\usepackage{textcomp}
\usepackage[]{subfigure}

\usepackage{booktabs}
\usepackage{array}
\usepackage{pgf}
\usepackage{graphicx}
\usepackage[absolute,overlay]{textpos}
\usepackage{calc}
\usepackage{amsmath}
\usepackage{amsfonts}
\usepackage{amssymb}
\usepackage{textcomp}
\usepackage{tikz}
\usepackage{gensymb}
\usepackage{color,soul}

\newcommand{\commentCWS}[1]%
{\textsf{\textcolor{blue}{#1$^{\mathrm{CWS}}$}}}
\usepackage[colorinlistoftodos]{todonotes} 

\newcommand{\figwidth}{0.9} 

\begin{document}

\begin{frontmatter}

\title{Vibratory Powder Feeding for Powder Bed Additive Manufacturing using Water and Gas Atomized Metal Powders}


\author[ubc]{C.~W.~Sinclair\corref{cor1}}
    \ead{chad.sinclair@ubc.ca}
\author[canmora]{R.~Edinger}
\author[ubc]{W.~Sparling}
\author[riotinto]{A. Molavi-Kakhki}
\author[riotinto]{C. Labrecque}

\cortext[cor1]{Corresponding Author,}

\address[ubc]{Department of Materials Engineering, The University of British Columbia, 309-6350 Stores Road, Vancouver, Canada} 
\address[canmora]{CANMORA TECH Inc., Richmond BC, Canada}
\address[riotinto]{Rio Tinto Metal Powders, 1625, route Marie-Victorin, Sorel-Tracy, QC J3R 1M6, Canada}

\begin{abstract}
Commercial powder bed fusion additive manufacturing systems use recoaters for the layer-by-layer distribution of powder. Despite the known limitations of recoaters, there has been relatively little work presented on the possible benefits of alternative powder delivery systems. Here, we show the use of a technology using simple vibration to control the powder flow for powder bed additive manufacturing. The capabilities of this approach are illustrated experimentally using two very different powders; a `conventional' gas atomized Ti-6Al-4V powder designed for electron beam additive manufacturing and a water atomized Fe-4wt\%Ni alloy used in powder metallurgy. Discrete element modelling is used to reveal the mechanisms controlling the dependence of feed rate on feeder process parameters and to investigate the potential strengths and limitations of this approach.  

\end{abstract}


\end{frontmatter}

\date{\today}


\section{Introduction} 

A feature common to all commercial powder-bed fusion additive manufacturing (PBF-AM) techniques is the use of powder recoaters for the layer-by-layer delivery of the powder onto the powder bed. While most aspects of process parameters in powder-bed based additive manufacturing have been explored in great detail (see e.g. \cite{Das2003PhysicalMetals,Korner2016AdditiveReview,Dowling2020AFusion,Mani2017MeasurementProcesses}), this mode of powder delivery has remained with few changes since the first commercial designs in laser and electron beam systems\cite{Mindt2016PowderInput,Yang2007MeteringTechniques,Hebert2016Viewpoint:Manufacturing}.  The basic elements of powder delivery using powder coating technology is largely the same across processes; powder is dosed according to the desired powder layer height from a hopper or reservoir, this powder then being leveled across the powder bed using a rake or, in some cases, a roller\cite{Hebert2016Viewpoint:Manufacturing}.  In the case of rakes, the edge can be hard (metal or ceramic edge) or soft (e.g. silcone or carbon fibre brush) with different advantages to both depending on the application. Optimizing rake shape to improve various characteristics of the deposited powder bed has also been investigated \cite{Haeri2017OptimisationSimulations}.   

It is generally recognized that the `quality' of parts built from powder bed based AM techniques relies on specific characteristics of the powder bed itself\cite{Meier2019CriticalManufacturing,Ali2018OnProcesses,Chen2019Powder-spreadingModeling,Haeri2017OptimisationSimulations,Mindt2016PowderInput}.  In particular, the layer thickness, roughness of the top surface and density of the powder bed have been shown to impact on part quality.  The process of re-coating can have a significant impact on the powder bed, not only through average properties like density and powder bed thickness but also more locally where the re-coater can introduce inhomogeneity in the powder bed. Defects in the powder bed can be introduced by the process of raking either due to damage to the rake itself or due to the selective pushing of larger particles ahead of the rake leading to a trail of reduced particle density parallel to the direction of rake travel\cite{Mindt2016PowderInput,Foster2020OpticalFusion}.  Additionally, the powder used in coaters must exhibit sufficient `flowability' or `spreadibility' so as to avoid the problems described above\cite{Meier2019ModelingSimulations}.  While the powder metallurgy industry has well established tests for flowability, these do not naturally translate to the recoating process and thus there has been significant interest in establishing PBF-AM specific `spreadability' tests \cite{Snow2019OnManufacturing}.  Insufficient powder spreadibility results in low rate of powder deposition, low powder bed density and very uneven powder distribution within the bed. For sufficiently low spreadability, jamming will occur leading to little or no spreading of the powder\cite{Snow2019OnManufacturing}. Highly `spreadable' powders are empirically known to consist of spherical particles with a narrow size distribution, these characteristics requiring gas atomization for PBF-AM powder production\cite{Hebert2016Viewpoint:Manufacturing,Anderson2018FeedstockDevelopment}.  

While gas atomized metal powders are standard in PBF-AM, there has been growing  interest in finding ways of using low `spreadability' water atomized metal powders owing to their significantly lower cost \cite{Hoeges2017AdditivePowders,Pinkerton2005DirectPowders,Fedina2020AManufacturing,Palousek2017ProcessingStudy,Letenneur2017LaserOptimization}. It has been shown to be possible to use water atomized metallic powders in selective laser powder bed melting \cite{Pinkerton2005DirectPowders,Letenneur2020LaserParts,Fedina2020AManufacturing}, though reports of success remain relatively few. No published results for water atomized powder used in selective electron beam melting have been reported to our knowledge. 

As noted above, many of the issues with powder feeding are a result of the interaction between the inherent powder characteristics \emph{and} the properties of the re-coater.  This begs the question of whether alternative powder feeding technologies could provide a route towards accommodating a much broader range of feed materials, including water atomized powder.  Taking a broader view, early work in additive manufacturing explored a variety of different powder delivery strategies (see e.g. \cite{Yang2007MeteringTechniques} for an early review).  Many of the explored techniques took inspiration from mature industries (e.g. pharmaceuticals, food, and automotive) where the delivery of granular mateƒrials is a vital step in manufacturing. Vibration is used to assist with powder delivery in these industries particularly when flow of the objects is a concern. Vibration during powder delivery can play a number of roles including inducing flow when the powder would not otherwise flow. Vibration can be used to break apart clusters of particles, to overcome powder adhesion and increase the free volume of the powder\cite{Yang2007MeteringTechniques,Tolochko2004LawsZone}. Providing kinetic energy to the granules can allow for a better exploration of configurational space and ultimately can allow for the avoidance of powder settling into a metastable configuration that blocks flow. It is unsurprising then, that vibration assisted powder feeding has been widely used in the powder metallurgy industry, for laser sintering and for laser cladding applications  \cite{Tolochko2004LawsZone} where irregularly shaped powders are common. 

In this work, we have explored the use of a vibratory powder feeder to distribute powder over the build plate with a specific eye towards application to non-conventional powder for additive manufacturing. We start by describing the powder feeding technology used followed by the materials tested in this preliminary study.  We characterize the feeder behaviour experimentally using two very different powders; the first a gas atomized Ti-6Al-4V powder conventionally used in electron beam additive manufacturing (see e.g. \cite{Sun2015ManipulationApplications}) and the second a commercial water atomized Fe-4wt\%Ni powder conventionally used in powder metallurgy.  While the former exhibits a highly spherical powder shape and narrow particle size, the latter exhibits a highly irregular powder shape and very wide size distribution.  Experiments combining feeding and electron beam based melting are also described and illustrative single layer squares are produced to illustrate the ability to feed and melt with the Fe-4wt\%Ni powder.  To better understand the relationship between feed rate and process parameters for this vibratory feeding approach, discrete element modelling (DEM) is performed.  The results of these simulations highlight the potential strengths and limitations. 

\section{Experimental Methodology:}

\begin{figure}[htbp]
\centering
\includegraphics[width=\figwidth\textwidth]{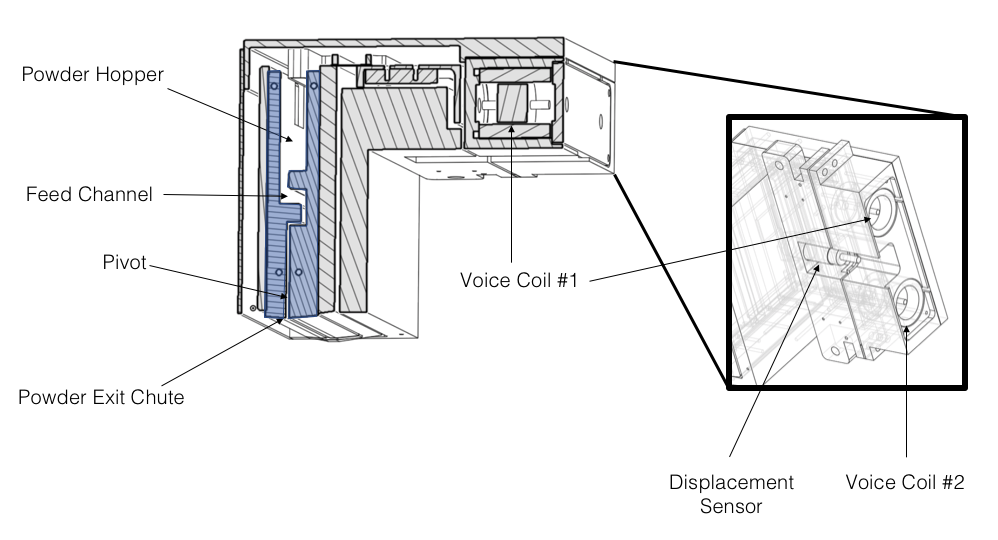}
\caption{Illustration of the Canmora TECH vibratory powder feeder used in this study.  The region in blue contains the powder.  Vibration induced by the voice coils causes powder to feed from the hopper, through the feed channel and out of the exit chute.  Displacement of the feeder induced by the voice coils is monitored using a separate displacement sensor.}
\label{fig:feeder}
\end{figure}

In this study, we have used of a recently developed technology (patent pending, Canmora TECH Inc.) for powder delivery in powder bed based additive manufacturing.  This technology uses vibration as the means of the controlled metering of metal powder onto a powder bed.  Figure \ref{fig:feeder} illustrates the relatively simple setup of the feeder. 

The key part of the feeder is highlighted in blue in figure \ref{fig:feeder}.  This inner portion of the feeder can be separated into three parts, a hopper that contains the powder to be fed, a feed channel that ensures powder does not feed when the feeder is off and an exit chute through which the powder is dispersed onto the powder bed. Two voice coils, electromagnetically shielded from the contents of the feeder, are attached as shown. The voice coils are actuated by means of a computer controller and amplifier so as to generate a sinusoidal displacement of specified amplitude (specified as a voltage) and frequency.  The displacement of the feeder is monitored independently by a separate displacement sensor mounted between the voice coils.  The inner portion of the feeder (blue) rotates on a bearing located close to the bottom exit chute of the feeder. 

The feeder was used in two separate setups. Feeding was performed both in air as well as within the high vacuum ( $< 5 \times 10^{-5}$ mbar) LEAM electron beam additive manufacturing facility at the University of British Columbia.  In both cases, the feed rate was calculated by placing a digital scale beneath the feeder and feeding for 10~s. The mass flow rate was then estimated based on the measured mass fed in that time.  In all cases feeding was performed starting from the feeder containing the same mass of powder so as to eliminate any variations induced by differences in feeder mass. It was noticed, however, that the feed rate did not significantly change with feeding as long as sufficient powder remained within the feeder so as to allow for continuous feeding.  In the case of the single layer melt experiments described below, a single powder layer of controlled thickness was deposited onto the build table by holding the feeder in a fixed position, controlling the feed rate of powder, and moving the build table under the feeder using a computer controlled x-y table.  

\begin{figure}[htbp]
\centering
\includegraphics[width=\figwidth\textwidth]{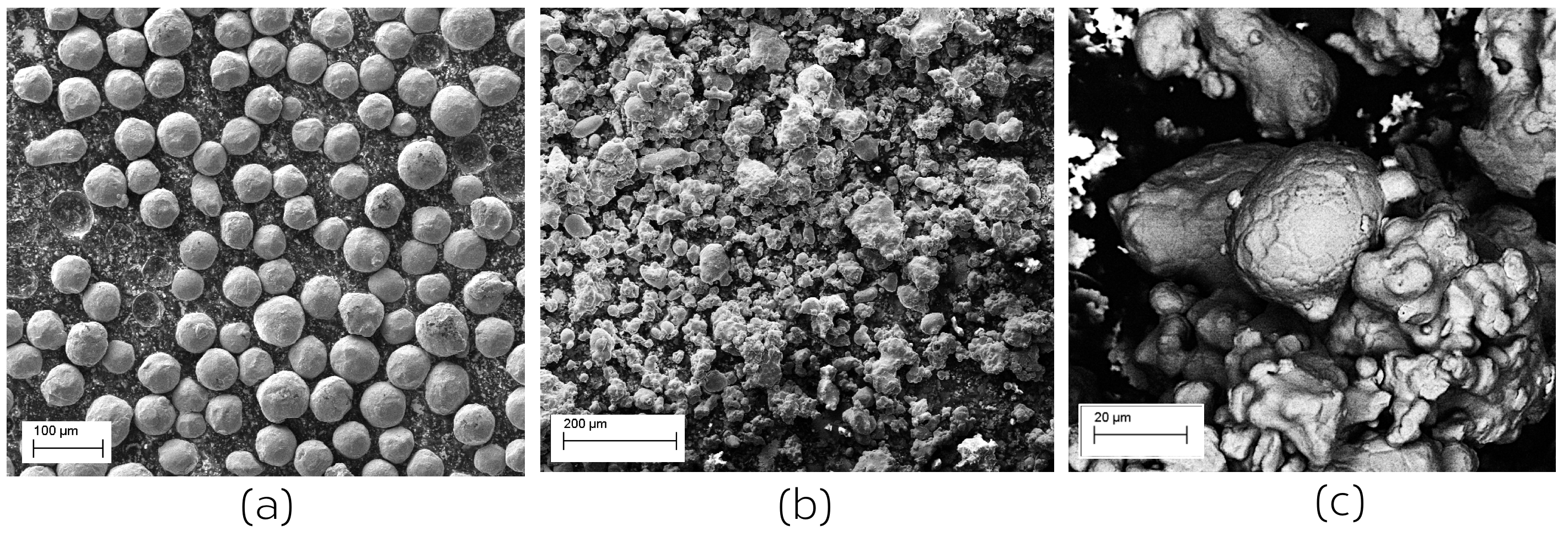}
\caption{a) Ti-6Al-4V gas atomized powder obtained from AP\&C b) Water atomized ATOMET 4801 Fe-4wt\%Ni powder provided by Rio Tinto Metal Powders c) magnified view of the ATOMET 4801 powder showing the irregular size and shape.}
\label{fig:powder}
\end{figure}

As mentioned above, two powders were used in this study.  The first was an AP\&C Ti-6Al-4V grade 23 with a size range of 45 - 106 $\mu$m and a D50 of 71 $\mu$m.  The characteristics of this powder have been discussed in previous publications (see e.g. \cite{Sun2015ManipulationApplications}).  The powder used had been recycled several times, the exact number not having been reported.  However, as can be seen in figure \ref{fig:powder} the morphology and size of the powder particles remains very regular and spherical with a relatively small proportion of satelite particles.

The second powder, ATOMET 4801, was provided by Rio Tinto Metal Powders.  This water atomized Fe-4wt\%Ni alloy is conventionally used in powder metallurgy applications.  This powder, illustrated in figure \ref{fig:powder}b, is very irregular in shape with a with a wide size distribution. Sieving produces results showing that 10wt\% of particles are larger than 150 $\mu$m in size (U.S. mesh +100), 62wt\% are larger than 45 $\mu$m in size (U.S. mesh +325) and 28wt\% are less than 45 $\mu$m in size (U.S. mesh -325).  The ATOMET powder was used in an un-recycled state.

While the two powders appear very differently in figure \ref{fig:powder}, they have similar apparent (tap) densities (2.47 g/cm$^3$ for Ti-6Al-4 \cite{APC} and 3.00 g/cm$^3$ for ATOMET \cite{RTMP} and the same flow rate (25 s/50g) as measured by Hall Flow meter following ASTM B213 \cite{APC,RTMP}.
 
Finally, the feeder has been used to perform simple single layer melt trials with the two powders. Here we focus only on the results for the Fe-4wt\%Ni powder with the aim of illustrating the potential for feeding and melting of this less conventional (for additive manufacturing) powder within electron beam additive manufacturing.  An AISI4340 steel build plate was used and a single layer of powder was delivered to the build plate by controlling the feed rate from the feeder and the translation of the build plate (on its x-y translation stage) beneath it.  This was done to achieve an approximate layer thickness of 0.2~mm.  Seven independent 15 $\times$ 15 mm squares were melted using a simple outer contour followed by infill by hatching.  Prior to depositing the powder layer, the build plate was heated to 940$^\circ$C to eliminate smoking.  For both hatching and contouring an electron beam accelerating voltage of 100 kV and a beam velocity of 3.3 mm/s was used.  The beam currents for each square varied from 0.8 to 2.0 mA in increments of 0.2 mA.  

\section{Experimental Results:}

\subsection{Frequency Controlled Feeding of Powders:}
\begin{figure}[htbp!]
\centering
\includegraphics[width=\figwidth\textwidth]{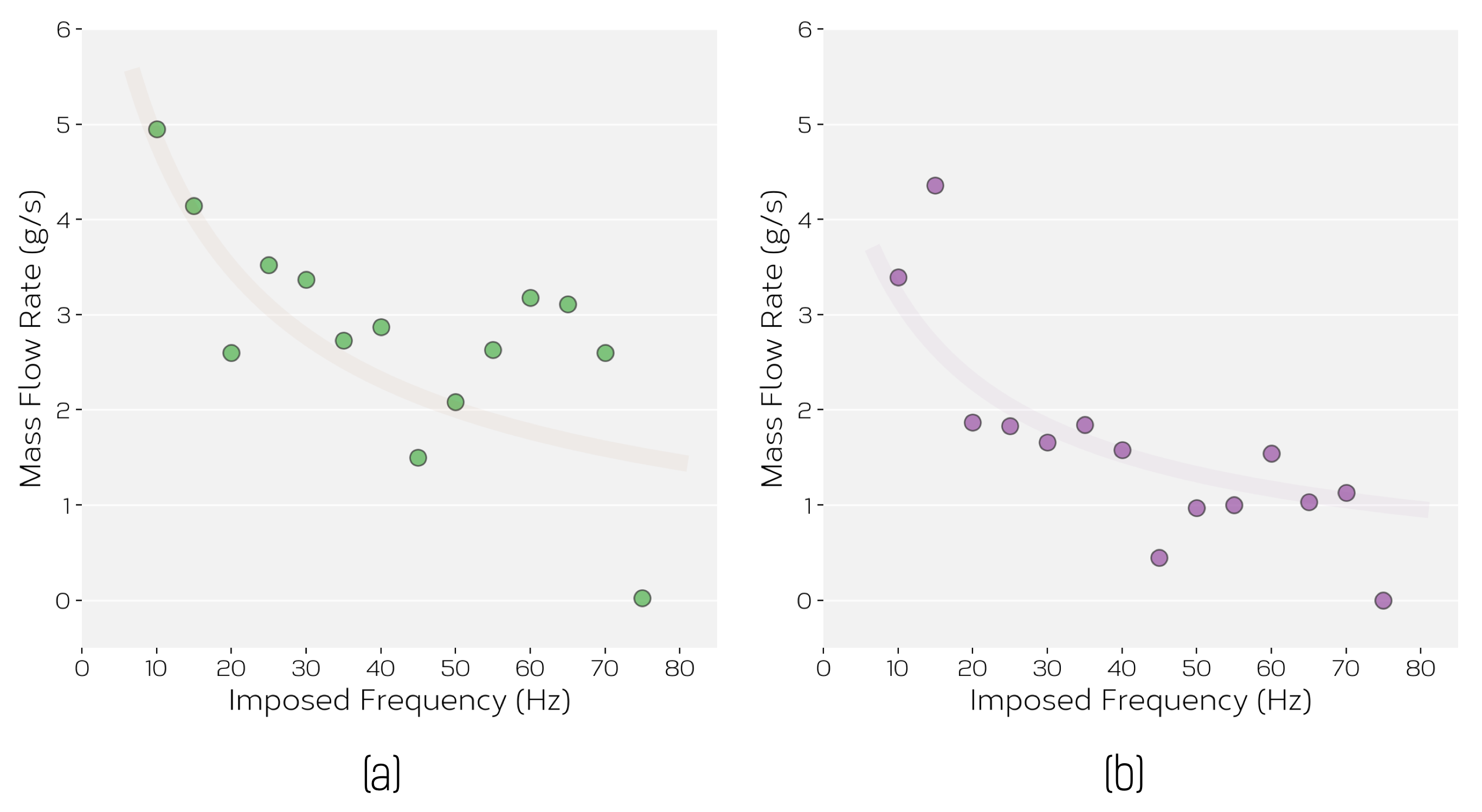}
\caption{The measured mass flow rate, estimated from the mass of powder delivered within 10~s, for a fixed voice coil displacement excitation (amplitude) and different voice coil frequencies for a) Ti-6Al-4V powder and b) ATOMET 4801.  The lines are intended only as a guide to the eye.}
\label{fig:feedrate_freq_exp}
\end{figure}

The construction of the feeder is such that powder only flows when the voice coil is activated, and even then only when the correct conditions are imposed.  We have evaluated this under two sets of conditions.  First, Figure \ref{fig:feedrate_freq_exp} shows the resulting feed rate for the two powders, fed in air, for different imposed frequencies of the voice coil at fixed driving amplitude (fixed imposed voltage maximum to voice coil).  While the results in figure \ref{fig:feedrate_freq_exp} show a general trend of increasing feed rate with reduced frequency, significant scatter exists within the data.  

\begin{figure}[htbp!]
\centering
\includegraphics[width=\figwidth\textwidth]{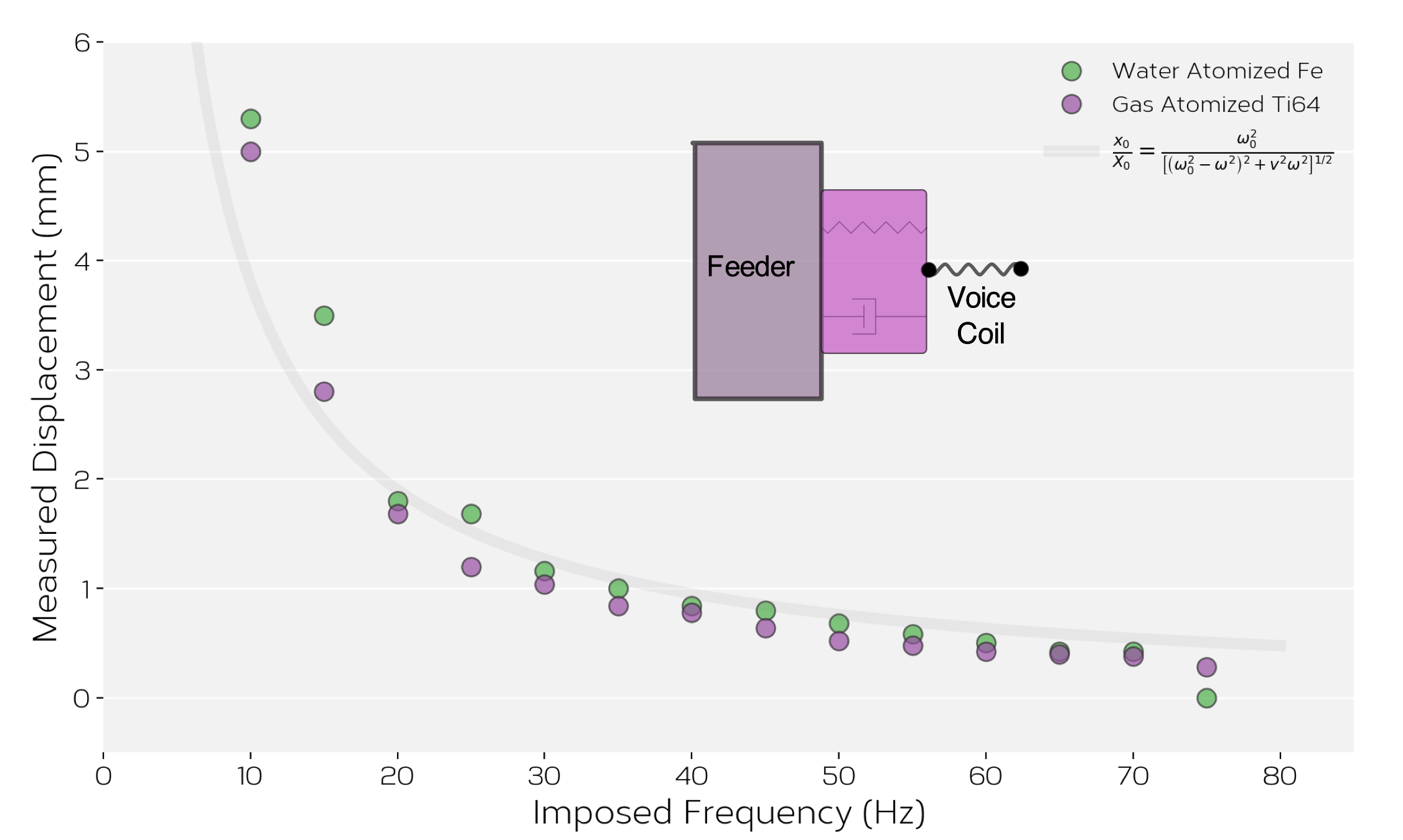}
\caption{Measured feeder displacement amplitude versus frequency for both Ti-6Al-4V and Fe-Ni powders when the imposed vibration amplitude of the voice coil was fixed as constant.  Also plotted as a solid line is the predicted behaviour assuming the feeder/voice coil system as a driven, damped harmonic oscillator.}
\label{fig:dispfreq}
\end{figure}

As the voice coil induced oscillations are imposed via its coupling to the feeder, control of the feed rate by varying voice coil frequency at fixed amplitude was found to be challenging.  In the setup used here it was found that it was not possible to independently control the feeder's vibration frequency and displacement as the system behaves as a driven, damped harmonic oscillator. This can be shown by means of figure \ref{fig:dispfreq}.  Here, the measured feeder displacement is shown as a function of frequency for the same conditions as in figure \ref{fig:feedrate_freq_exp}, the same color scheme being used to indicate the results from the two different powders.  Here, although the amplitude (voltage) imposed on the voice coil was fixed to be constant, the measured feeder displacement was found to increase strongly with decreasing frequency, independent of the type of powder used.  This dependence is exactly the one that would be expected for a driven, overdamped harmonic oscillator. A system of mass $m$ driven through a spring (constant $k$) and dashpot (damping coefficient $\nu$) by a source with amplitude $X_0$ and frequency $\omega$ will respond with an amplitude $x_0$ such that,

\begin{equation}
   \frac{x_{0}}{X_{0}}=\frac{\omega_{0}^{2}}{\left[\left(\omega_{0}^{2}-\omega^{2}\right)^{2}+v^{2} \omega^{2}\right]^{1 / 2}}
\end{equation}

where $\omega_0 = \sqrt{k/m}$ is the natural frequency of the system.  This relationship is plotted as a line on figure \ref{fig:dispfreq} with assumed values for $X_0$ and $\omega_0$.  As can be seen this explains this data well under the condition that the feeder contains the same mass ($m$) of the two powders as noted above.  This result also implies that when $\omega = \omega_0$ resonance will occur.  Evidence of this was observed experimentally as, at certain combinations of imposed voice coil frequence/amplitude, the measured displacement signal was found to change from a well controlled, regular sinusoidal variation, to a more chaotic pattern indicative of $\omega \rightarrow \omega_0$.  

\subsection{Displacement Controlled Feeding:}

\begin{figure}[htbp!]
\centering
\includegraphics[width=\figwidth\textwidth]{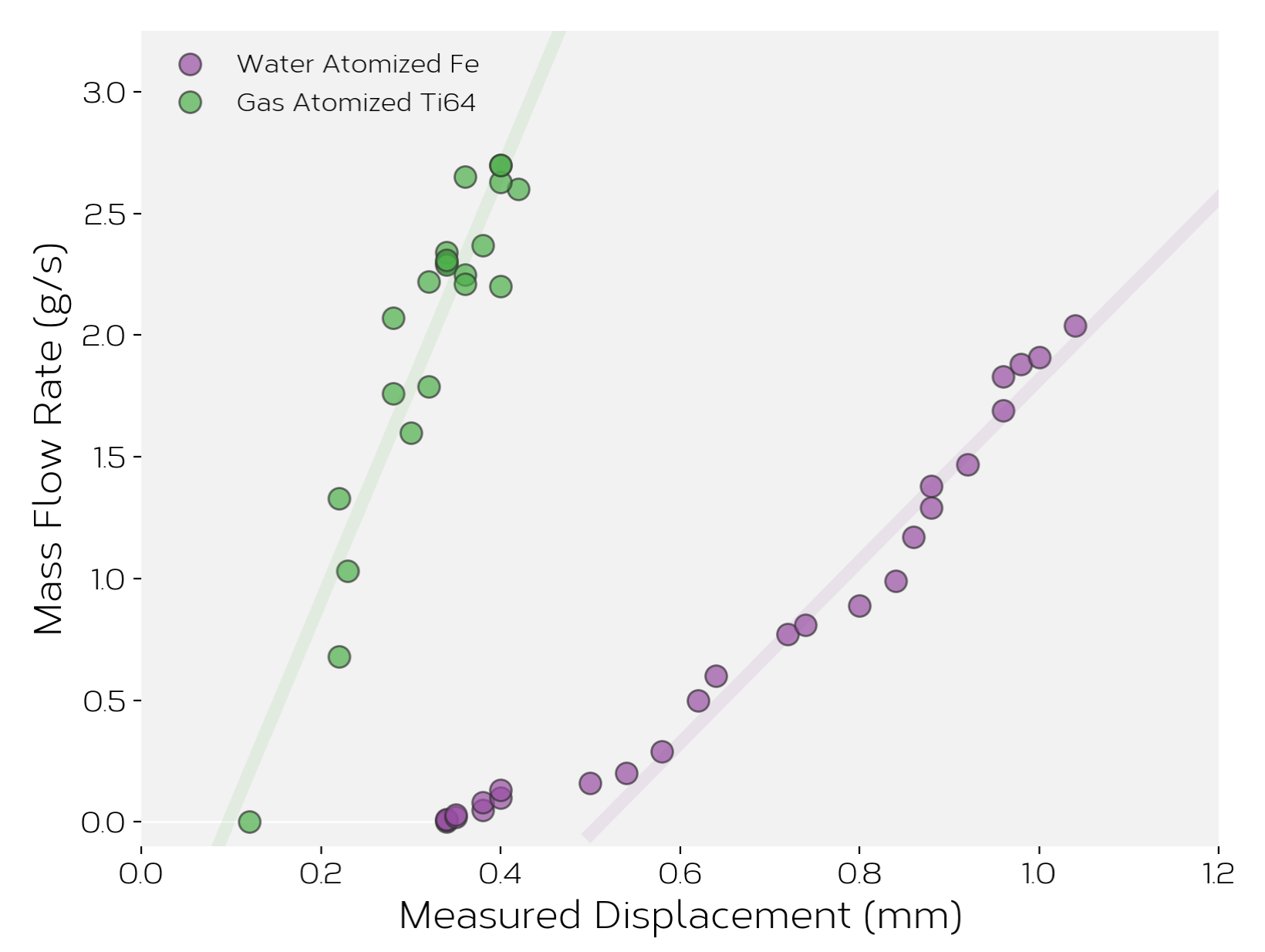}
\caption{The mass flow rate as a function of measured feeder displacement amplitude for both gas atomized Ti-6Al-4V and water atomized ATOMET 4801 powders. The lines are only intended as a guide to the eye.}   
\label{fig:flowdisp}
\end{figure}

The results in figure \ref{fig:dispfreq} show only a small dependence of feeder displacement on imposed frequencies for frequencies greater than $\sim$40 Hz.  Thus, further experiments and development were focused onto controlling feed rate by fixing imposed frequency, at $\omega > 40$ Hz and varying the displacement of the feeder using the measured displacement from the displacement sensor.  Figure \ref{fig:flowdisp} shows the resulting feed rates measured in air for the two powders when the voice coil frequency was fixed to 57 Hz and the imposed amplitude varied to give the (fixed) feeder displacements shown. Under these conditions, one can see that the mass flow rate varies nearly linearly with the feeder displacement above a lower threshold displacement for both powders.  The mass flow rate for the Ti-6Al-4V powder is higher for all tested displacements, but it is possible to obtain the same stable flow rate for both the highly spherical Ti-6Al-4V and highly irregular ATOMET 4801 powder by the judicious selection of the feeder displacement.  While figure \ref{fig:flowdisp} was generated from experiments performed in air, the same experiments were repeated for feeding within the vacuum of the LEAM chamber.  No significant differences from the feeding in air and vacuum were observed.

Figure \ref{fig:flowdisp} shows that the feeder operated under displacement control at fixed frequency is not only able to feed conventional gas atomized powder but also the highly irregular Fe-Ni alloy powder with no further modifications.  While the flow rate is significantly lower than that of the Ti-6Al-4V powder for a given feeder amplitude, it is still possible to achieve the same flow rate for the two powders by adjusting the feeder's displacement amplitude. 

\subsection{Single Layer Melt Trials on ATOMET 4801 Water Atomized Powder}

\begin{figure}[htbp!]
\centering
\includegraphics[width=\figwidth\textwidth]{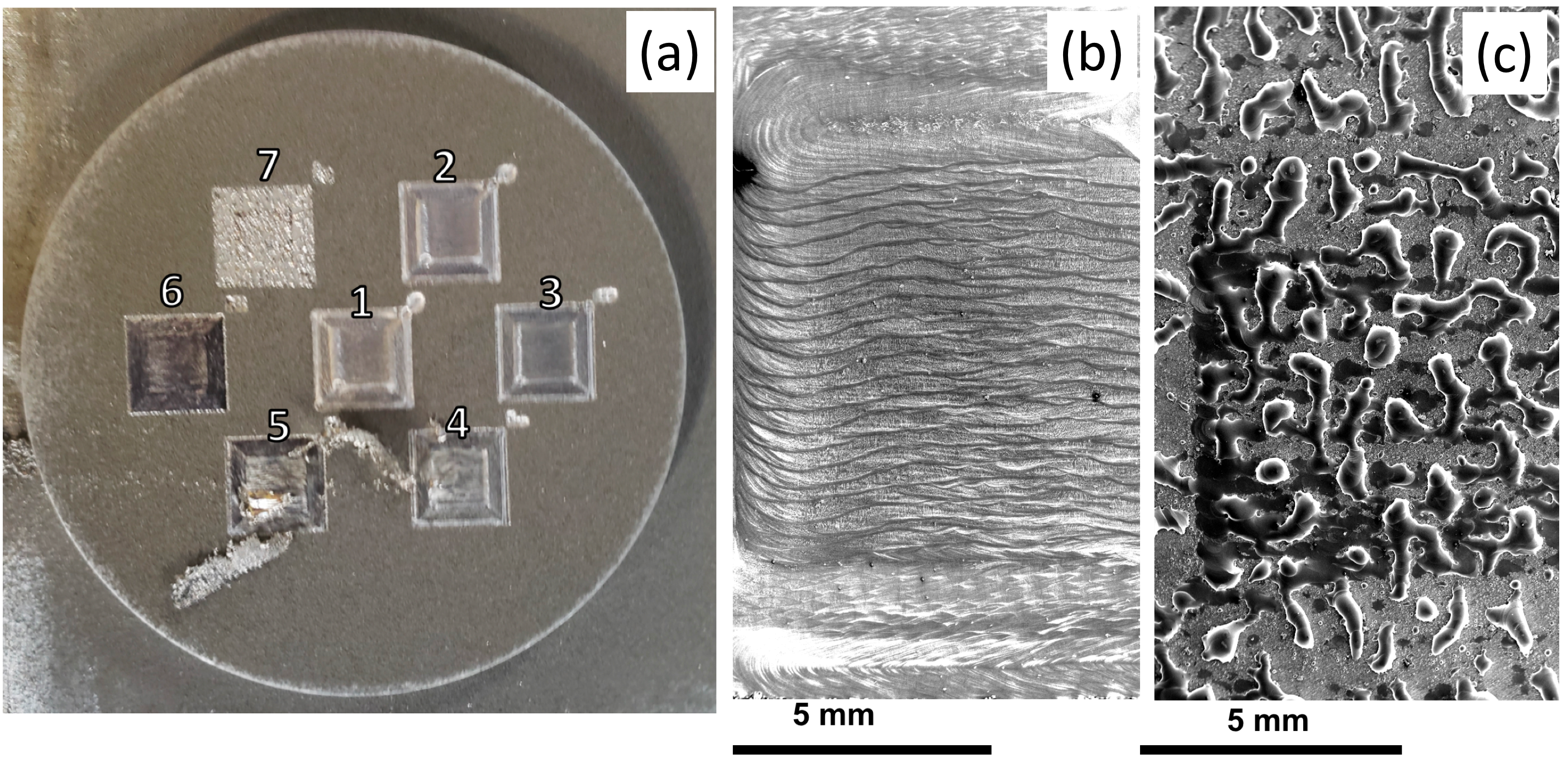}
\caption{a) Top down view of single layer melt trials for different electron beam conditions using the ATOMET 4801 powder.  The experiments were performed with an accelerating voltage of 100 kV, beam velocity of 3.3 mm/s (hatching) and beam currents of ranging from 2.0 mA to 0.8 mA in decrements of 0.2 mA for cases 1-7 b) Low magnification secondary electron image showing the hatched region for a) case 1 (2.0 mA) and b) case 7 (0.8 mA). All but case 7 showed nearly identical melt tracks to case 1.  Case 7 was the only to show clear evidence of under melting.} 
\label{fig:macromelt}
\end{figure}

\begin{figure}[htbp]
\centering
\includegraphics[width=\figwidth\textwidth]{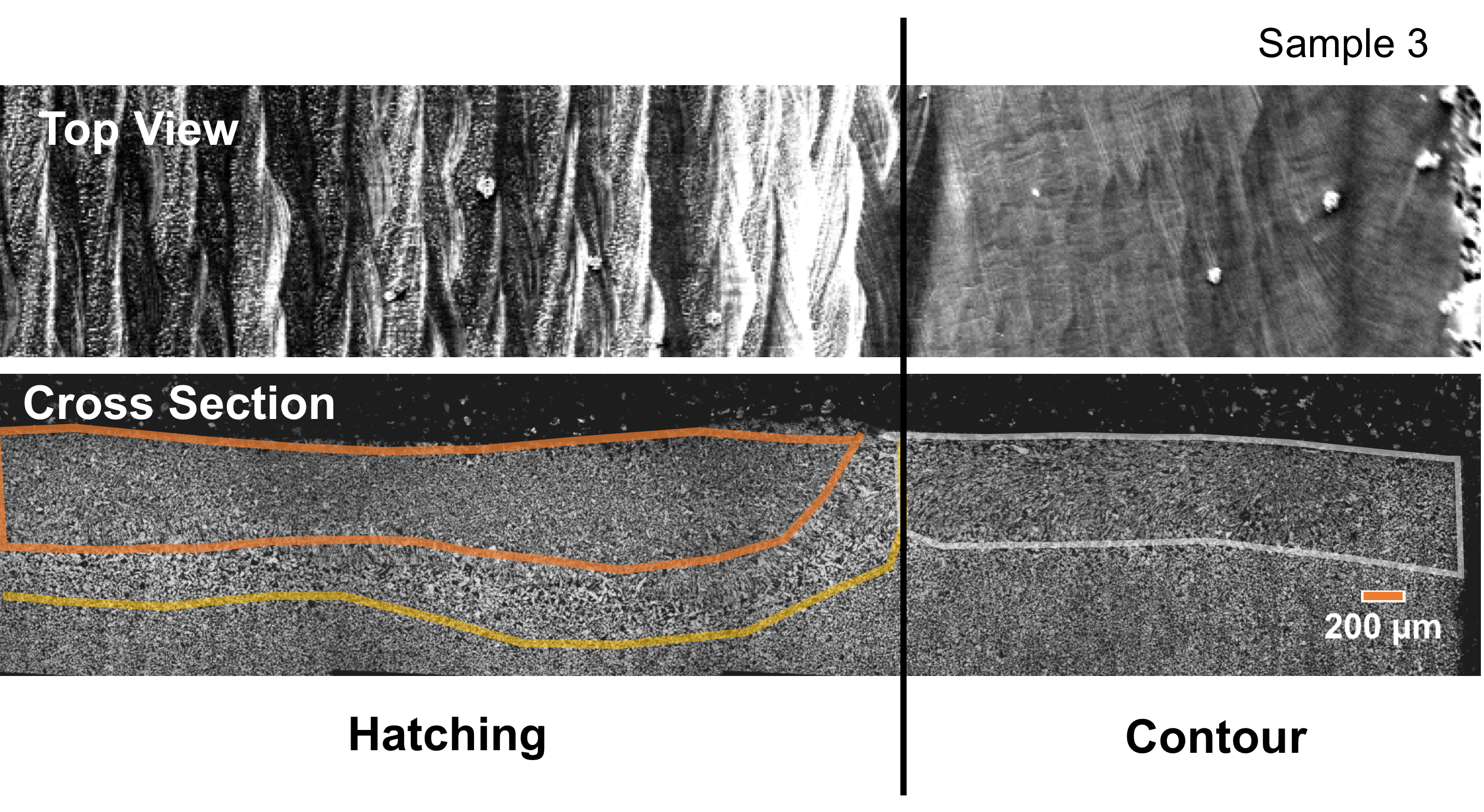}
\caption{A higher magnification view (top down) of case 1 from Figure \ref{fig:macromelt} along with secondary electron images taken through the cross section.  Here, one can see significant mixing with the underlying base plate and a clear heat affected zone in the region where hatching was performed.}
\label{fig:micromelt}
\end{figure}

With this result in mind, single layer melt trials were conducted to illustrate the potential for feeding of the Fe-4wt\%Ni powder. Figure \ref{fig:macromelt}a shows a macroscopic, top down view of the 7 squares produced, square 1 being produced with the highest beam current and square 7 with the lowest.  Macroscopically, it can be seen that all squares appear similar with only square 7 showing a significant difference from the rest.  Figures \ref{fig:macromelt}b and c show higher magnification top down views taken via scanning electron microscopy from (b) square 3  and (c) square 7. The appearance of the melt tracks in figure \ref{fig:macromelt}b are representative of the melt tracks observed for all other squares, save square 7.  Square 7 (\ref{fig:macromelt}c), on the other hand showed clear evidence of balling due to insufficient input power for consistent melting. Figure \ref{fig:micromelt} shows a cross-section through square 3 showing the uniformity of the melt depth and the lack of evidence for large scale defects (e.g. lack of fusion or `key hole' formation).  This was consistent for all squares, save square 7.    

\section{Interpreting Feeder Behaviour using DEM simulations}

\begin{figure}[htbp]
\centering
\includegraphics[width=\figwidth\textwidth]{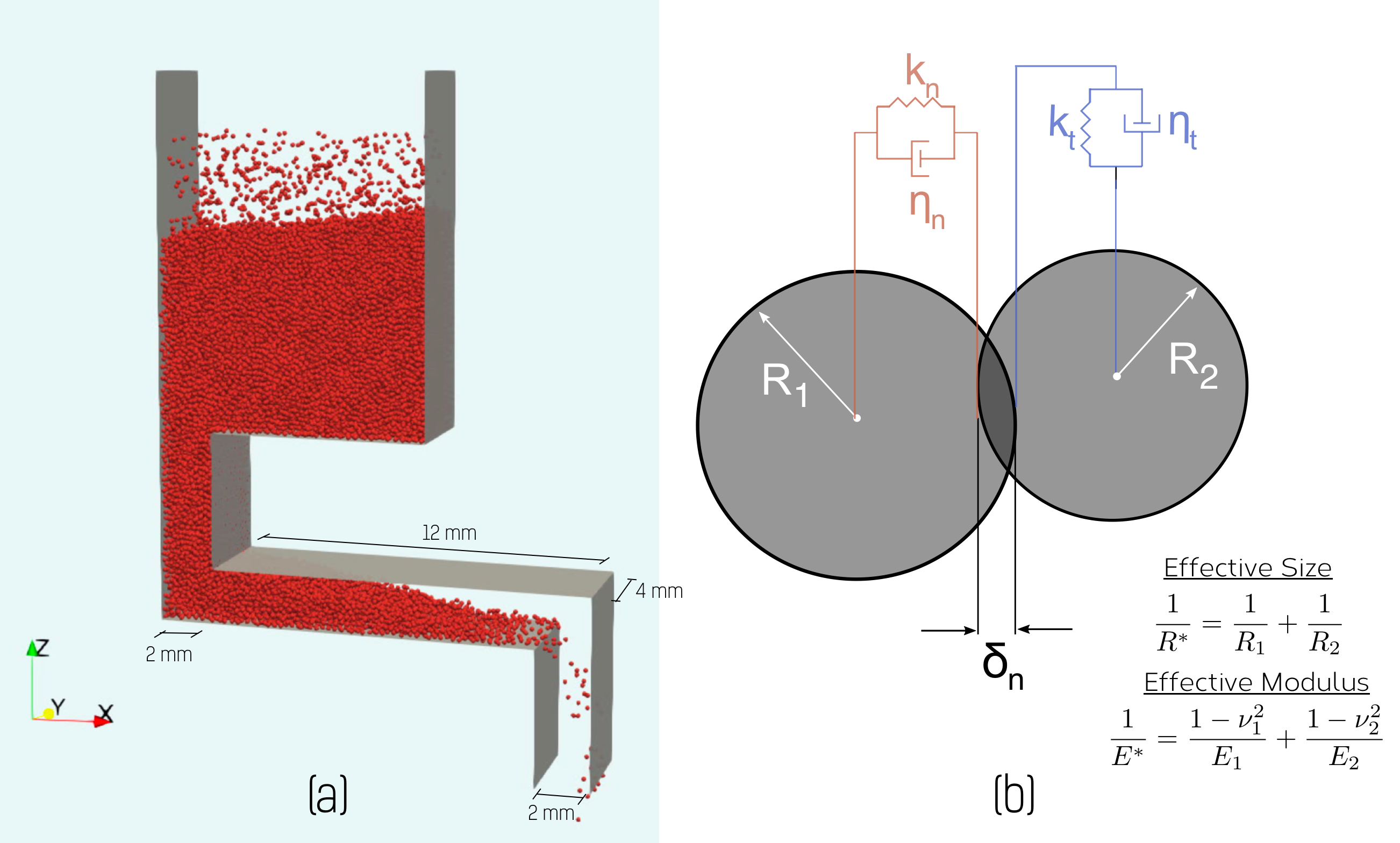}
\caption{a) Illustration of the model used for simulating feeder behaviour.  The model is periodic in the y-direction and the surfaces shown were taken to have the same contact behaviour as the particles (red).  b) Illustration of the basic elements of the Hertzian contact model used in simulating the behaviour of powder particles.}
\label{fig:DEM}
\end{figure}

In order to understand the behaviour exhibited by the feeder described above, particularly the response exhibited in figure \ref{fig:flowdisp}, we have used discrete element (DEM) simulations.  This follows on a large body of work using DEM simulations to understand powder flow in additive manufacturing (see e.g. \cite{Meier2019CriticalManufacturing, Meier2019ModelingSimulations,Haeri2017OptimisationSimulations,Herbold2015SimulationMethod,Parteli2013DEMManufacturing,Chen2019Powder-spreadingModeling}  Rather than attempt to quantitatively predict the experimental results, the aim of the simulations performed here was to reveal the mechanisms underlying controlling mechanisms. The geometry of the simplified feeder setup used in the models is shown in figure \ref{fig:DEM}a.  The model consists of hopper, feed channel and exit chute but is simplified, compared to the actual feeder, so as to be periodic in the y-direction.  This allowed for a significant speedup in simulation time (owing to the need to track fewer particles) though at the expense of neglecting powder-feeder wall interactions in the y-direction. Simulations were performed using the LIGGGHTs DEM package \cite{Kloss2012ModelsCFD-DEM} built on of the LAMMPS simulation platform \cite{Plimpton1995FastDynamics}.  Simulations used a Hertz-Mindlin no-slip contact model with a simplified JKR cohesion model to simulate particle-particle adhesion. Briefly, the equations of motion for each particle $i$ of mass $m_i$ are solved according to,

\begin{equation}
    m_i = m_i g + \sum_{j \neq i}^N F_{n,j} + F_{t,j} + F_{coh,j}
\end{equation}

where $g$ is acceleration to due gravity and the normal force is calculated for particles of a single type and size (radius, $R$) as,

\begin{equation}
    F_{n,j} = \frac{4}{3}\frac{E}{1-\nu^2}R^{1/2}\delta_n^{3/2}-\eta_n v_n
\end{equation}

and the tangential force is,

\begin{equation}
    F_{t,j} = \frac{8E}{2\left(2-\nu\right)\left(1+\nu\right)}R^{1/2}\delta_n^1/2\delta_t - \eta_t v_t
\end{equation}

The parameters $\delta_n$ and $\delta_t$ correspond to the normal and tangential overlap of contacting surfaces.  The damping coefficients $\eta_n$ and $\eta_t$ used above follow the dependence on the coefficient of restitution as described in \cite{Tsuji1992LagrangianPipe}.

Rather than use the simplified JKR model implemented in LIGGGHTS, we modified this model to use it in its conventional form,

\begin{equation}
    F_{coh,j} = - \gamma_{coh}E^{1/2}R^{3/4}\delta_n^{3/4}
\end{equation}

where $\gamma_{coh}$ here is a material parameter that measures the cohesive strength of the interaction (see e.g. \cite{Meier2019ModelingSimulations}).

The material parameters used here are defined in Table \ref{tab:DEMparams}, these having been strongly inspired by the experimental and simulation work on Ti-6Al-4V powders as reported in the work of Meier \emph{et al.} \cite{Meier2019ModelingSimulations}. An inherent limitation of DEM simulations applied to `stiff' materials, is the short time scale required to resolve particle-particle interactions \cite{Lommen2014DEMMaterial,Otsubo2017EmpiricalSimulations}.  Taking properties consistent with titanium would require one to use a time step of $<$10~ns, this requiring 100 million time-steps to simulate 1 second of feeder operation. To overcome this timescale limitation, we follow the convention in the field using a `soft particle' approximation.  Rather than using the actual modulus for Ti, the particles are assumed to have a much reduced elastic modulus, this leading to a much longer interaction time during particle contact.  It has been shown that for conditions similar to those used here the effect on predictions is small (see e.g. \cite{Lommen2014DEMMaterial,Meier2019ModelingSimulations}).  Time-steps were selected to be sufficiently small so as to satisfy the Rayleigh criterion \cite{Otsubo2017EmpiricalSimulations}. Most simulations shown here were performed with mono-sized, spherical particles.  A relatively large particle radius of 120~$\mu$m was used to again reduce the total number of particles simulated and thus the total required simulation time.  Tests performed for particle radii down to 50~$\mu$m showed no appreciable difference in the qualitative trends illustrated below.

\begin{table}[htbp]
 \centering
 \begin{tabular}{cc}
  \toprule
  \textbf{Property} & \textbf{Value}  \\
  \midrule
  Density, $\rho$ & 4.5 g/cm$^3$\\
  Poisson's Ratio, $\nu$ & 0.3 \\
  Young's Modulus,$E$ & 0.1 GPa \\
  Coefficient of Restitution, $e$ & 0.3 \\
  Cohesion Parameter, $\gamma_{coh}$ &  1.4 $\times 10^{-5} J^{1/2}/m$ \\
  \bottomrule  
 \end{tabular}
 \caption{Material Parameters used in DEM model}
 \label{tab:DEMparams}
\end{table}

\begin{figure}[htbp]
\centering
\includegraphics[width=\figwidth\textwidth]{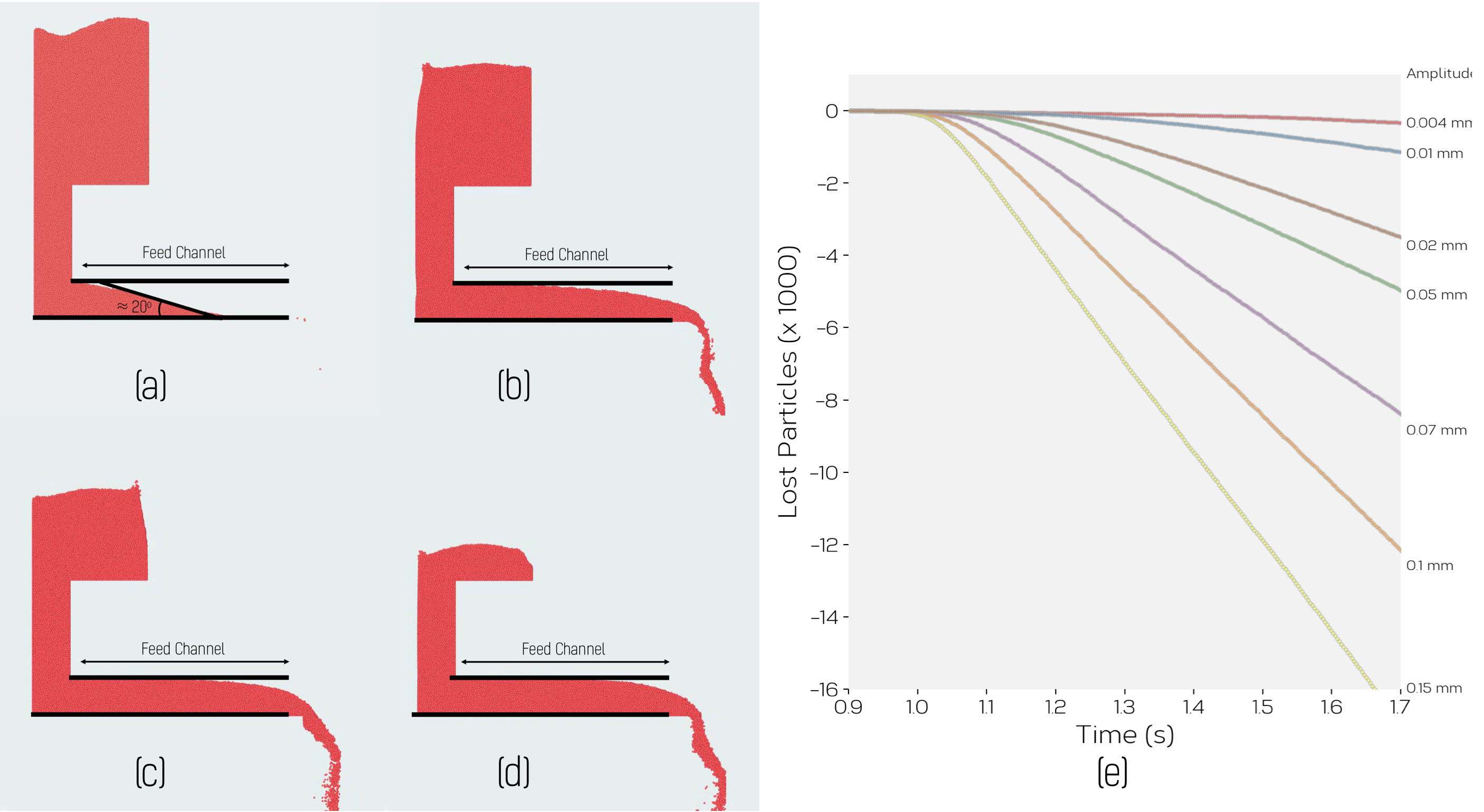}
\caption{a) - d) Illustration of progressive feeding of powder from the simulated feeder. Simulation results for particle radius of 50$\mu$m, vibration frequency and amplitude of 40 Hz and 0.25 mm. a) represents the initial configuration prior to feeding. e) The number of particles fed as a function of time for different displacement amplitudes of the feeder.  In this case the particle size was 120 $\mu$m and the vibration frequency was 100 Hz.  As one can see, the feed rate remains constant, for a given displacement amplitude and frequency, over the entire range of feeding.}
\label{fig:feedingdem}
\end{figure}

Simulations involved two steps.  First, particles were inserted into a region at the top of the feeder via a rain model \cite{Meakin1987RestructuringDeposition} and allowed fall into the hopper  until the hopper was nearly filled. The powder naturally flowed into the feed channel establishing a stable distribution within it.  This is illustrated in figure \ref{fig:feedingdem}a which shows the simulation setup at the end of the feeding step where the characteristic angle adopted by the powder within the feed channel, related via the assumed contact mechanics and cohesion model to the angle of repose for the powder, can be seen.  This observation immediately explains why powder does not continuously feed through the feeder without vibration; so long as the feed channel is long enough to contain this characteristic powder angle then the powder will not flow so long as the feeder is not vibrating.

Each simulation of feeding started from the same initial equilibrium distribution of powder within the feeder (figure \ref{fig:feedingdem}a).  To simulate vibration in this case, the entire feeder was subjected to a sinusoidal horizontal displacement in the $x$-direction, $\Delta x = A \sin{\left(\omega t\right)}$, with amplitude $A$ and frequency $\omega$. 

Upon initiating vibration, powder begins to flow through the feeding channel and out the exit chute.  This is illustrated for one condition in the sequence of images shown in figures  \ref{fig:feedingdem}a-d.  The feed rate in this case was quantified by measuring the number of particles to exit from the bottom of the chute as a function of time.  This is shown for different values of the vibration amplitude $A$ at a fixed value of frequency (100 Hz) in figure \ref{fig:feedingdem}e.  Figures  \ref{fig:feedingdem}a-d show flow for an amplitude A = 0.25 mm and frequency of 40 Hz.  As one can see, following an initial period required for powder to flow from its initial position to the exit chute, a constant rate of powder flow (linear relationship between number of particles lost and time) is established and maintained until nearly all powder has been removed from the feeder. The constant rate of feeding, regardless of the mass of powder in the feeder means that we can assess the feed rate as a function of amplitude and frequency without having to be concerned with the time dependence of the feed rate.  This also corresponds well with our experimental observations which also showed that the feed rate was not strongly sensitive to the mass of powder remaining in the feeder.

\begin{figure}[htbp]
\centering
\includegraphics[width=\figwidth\textwidth]{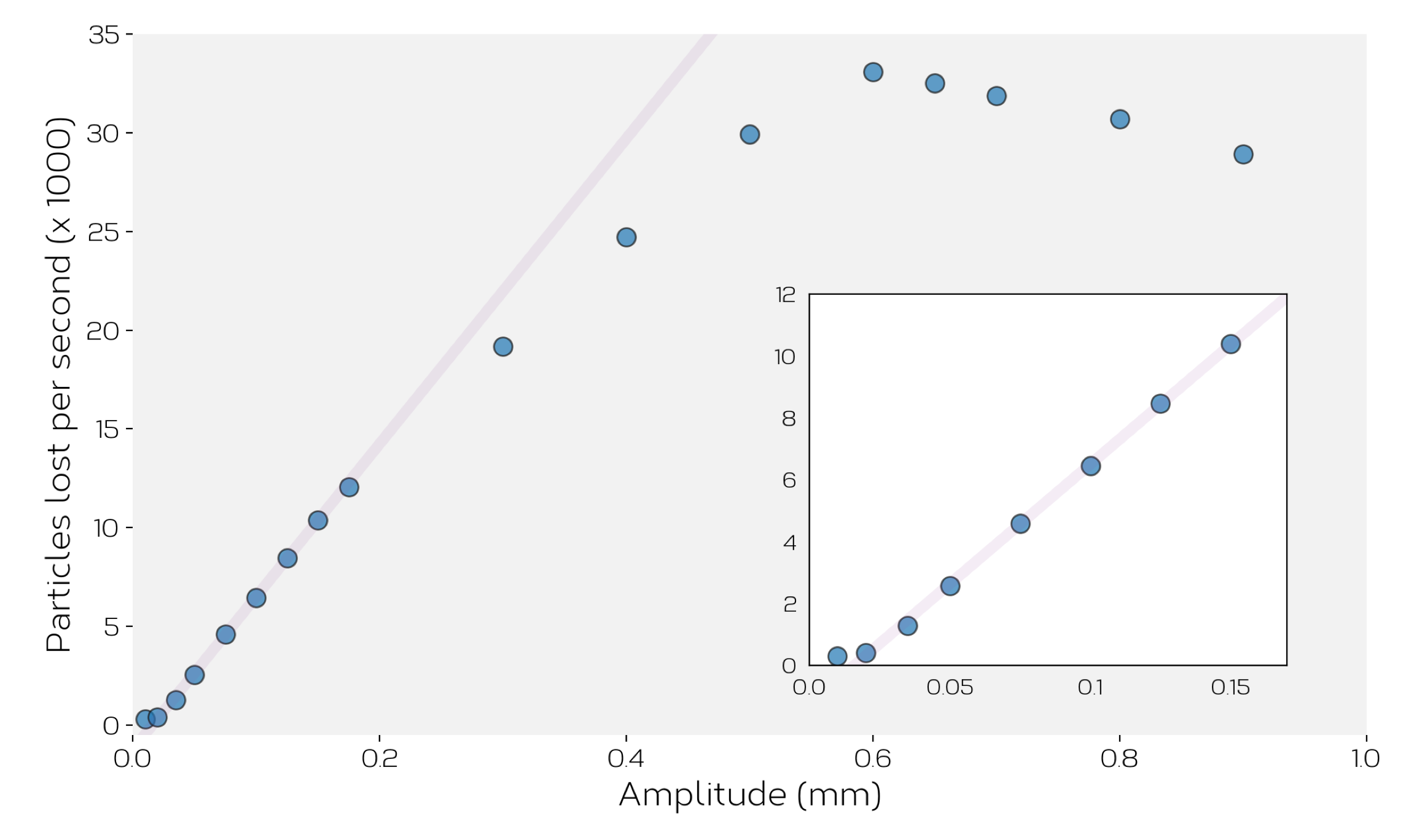}
\caption{DEM model prediction of steady-state particle feed rate as a function of amplitude at a fixed imposed frequency of 40 Hz. Inset shows the low amplitude portion of the plot magnified to show that feeding stops at a finite displacement amplitude.  The line is shown as a guide to the eye. }
\label{fig:feedamp}
\end{figure}

Figure \ref{fig:feedamp} shows the predicted feed rate as a function of vibration amplitude, $A$, for an imposed frequency of 40 Hz. One can see in this plot that for amplitudes between 0 and $\sim$0.4 mm the linear relationship observed experimentally in figure \ref{fig:flowdisp} is well reproduced by the simulations, as is the cut-off of flow below a certain imposed amplitude.  Below this amplitude the feeder oscillates but no powder is seen to leave the feeder.  

At amplitudes $A \gtrapprox 0.4$ mm one sees that the feed rate deviates away from linear, and for amplitudes above $A \gtrapprox 0.6$ the feed rate is seen to decrease.

\begin{figure}[htbp]
\centering
\includegraphics[width=\figwidth\textwidth]{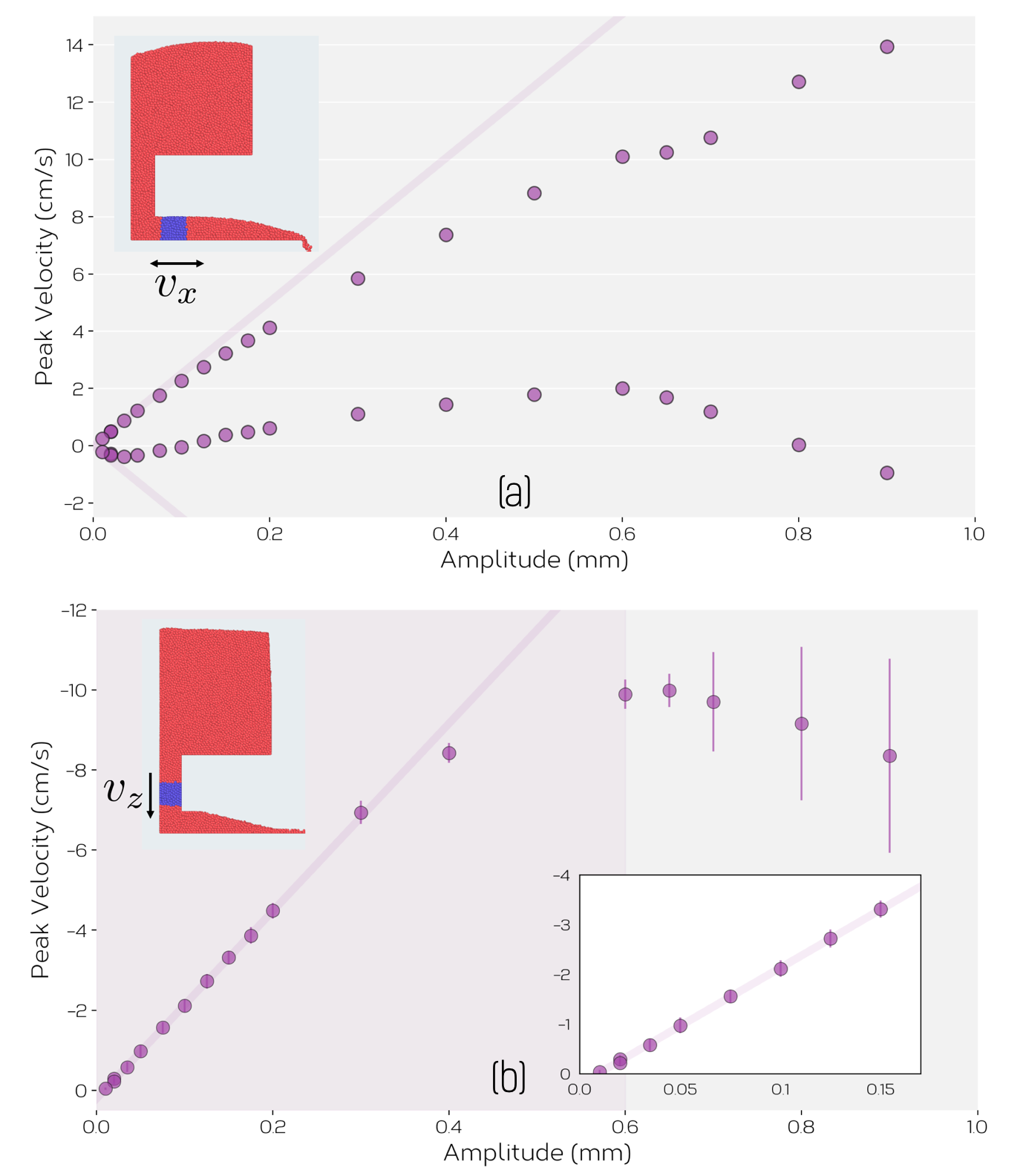}
\caption{a) Average horizontal velocity of particles in the region of the feed channel illustrated in blue b) Average vertical velocity of particles in the region of the feed chute illustrated in blue.  The inset in (b) shows a magnified view at small displacement amplitudes.}
\label{fig:vel}
\end{figure}

We can understand the behaviour in figure \ref{fig:feedamp} better by looking at how powder flows through the system (figure \ref{fig:vel}). In particular we focus on the vertical and horizontal velocities of the powder at different locations in the simulated feeder.  As the feeder moves back and forth in the $x$-direction, the powder also moves in a cyclic fashion meaning that we can identify a maximum and minimum velocity of particles in $x$-direction. This is shown in  figure \ref{fig:vel}a for particles in the blue shaded region of the feed channel. When the feeder displacement amplitude is small, the powder displacements are small and in an opposite sense on opposite sides of the (forward and backward) strokes.  The lines drawn on figure \ref{fig:vel}a show the maximum forward and backward velocities that would be expected of the powder if it simply followed the motion of the feeder.  It can be seen that for small feeder displacements the powder closely follows the motion of the feeder, simply moving back and forth as the feeder oscillates.  This creates no net flow of the powder through the feed channel.  As displacement amplitude is increased, the forward motion of the powder continues to follow that of the feeder (in the positive direction, towards the chute) but the reverse direction of flow saturates, and eventually starts to increase, eventually the flow of powder continuing to be in the positive $x$-direction though the direction of feeder is in the negative $x$-direction.  At small displacements, the forward stroke of the feeder compresses the powder against the force induced between the particles and the top and bottom of the feed channel walls. On the reverse stroke the powder recovers and flows backward to recover its original position.  This, apparent `compressibility' of the powder saturates as the force of friction does not continue to increase with larger displacements. Rather, at larger displacement amplitudes, the inertia induced by the feeder in the forward stroke is sufficient to keep the powder flowing in the forward direction for the time that the feeder direction is reversed. Thus, at small amplitudes the forward and backward motion of the powder cancel leading to no net flow.  As amplitude is increased, the net flow follows roughly that given by the motion during the forward stroke of the feeder as little motion occurs in the reverse stroke.  Thus, under these conditions, feeding can be described as a form of `ratcheting' with the flow being induced largely be the inelastic translation of powder induced in the forward portion of the stroke.

\begin{figure}[htbp]
\centering
\includegraphics[width=\figwidth\textwidth]{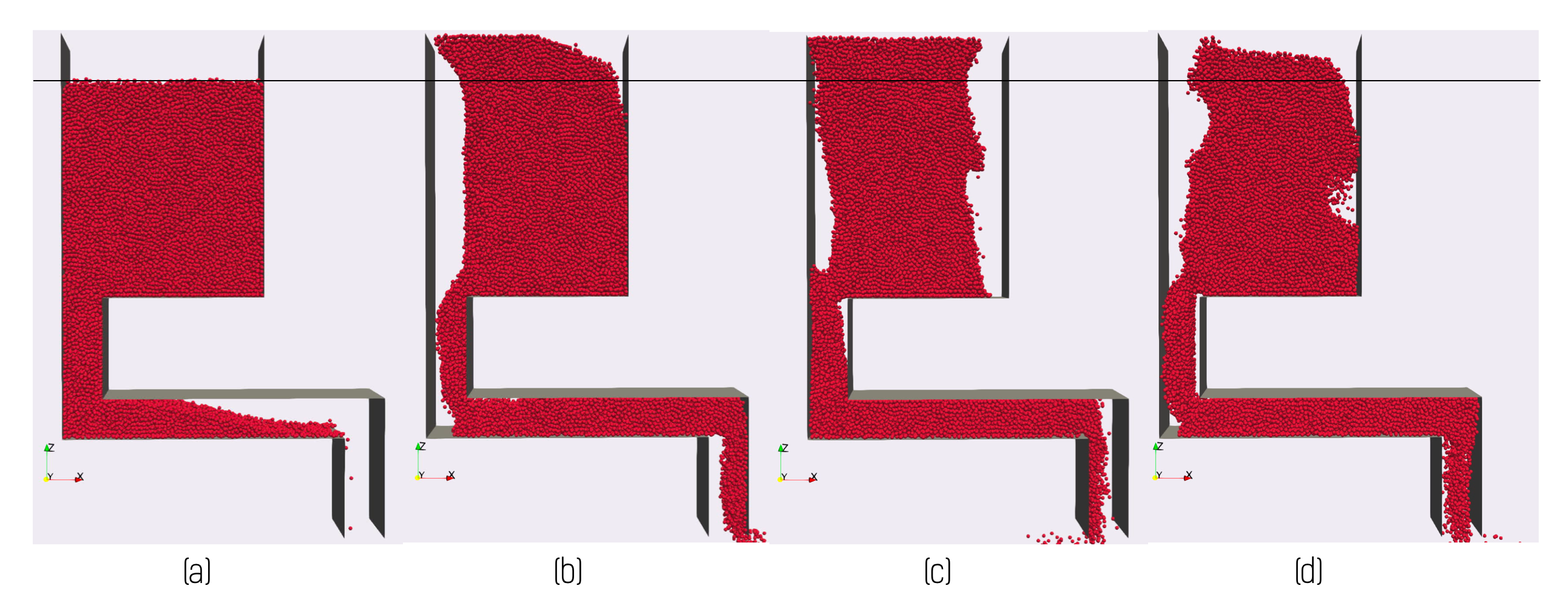}
\caption{Series of snapshots from a DEM simulation performed with a frequency of 40 Hz and amplitude $A$ = 0.8 mm showing the significant upward displacement of the powder in the hopper.  This upward motion results in a decreased delivery of powder into the feed channel and thus a reduction in the feed rate.  }
\label{fig:unstable}
\end{figure}

This explanation describes the behaviour illustrated in figure \ref{fig:flowdisp} but as can be seen in figure \ref{fig:feedamp} the rate of flow from the feeder reduces beyond a critical amplitude (here $\sim$ 0.6~mm).  To understand this behaviour it is instructive to look at the vertical flow of powder from the hopper into the feed channel as shown in figure \ref{fig:vel}b.  Here, one can see that the flow of powder from the hopper into the feed channel increases linearly with amplitude for amplitudes $A \lessapprox 0.4$ mm with a cutoff at small amplitudes, consistent with the observations in figure \ref{fig:vel}a.  At high amplitudes, $A \gtrapprox 0.6$ mm, the feed rate from hopper to chute is seen to decrease.  The reasons for this can be seen graphically in figure \ref{fig:unstable}.  Here it can be seen that high displacement amplitudes result in significant upward motion of the powder in the hopper (compare to the relatively quiescent behaviour of the powder in the hopper shown in figure \ref{fig:feedingdem}).  This upward motion is a result of the high velocity of the feeder and an inability for the powder to relax and follow the horizontal translation of the feeder. In this case, the vertical motion of the powder reduces the amount of powder flowing into the feed channel and thus the feed rate.  This illustrates the need for careful consideration of the operation window of the feeder to ensure that feeding is conducted in a regime where powder flow remains stable both within the hopper and feed channel. This sets a limit on the maximum feed rate that can be obtained under displacement controlled feeding.  

\begin{figure}[htbp]
\centering
\includegraphics[width=\figwidth\textwidth]{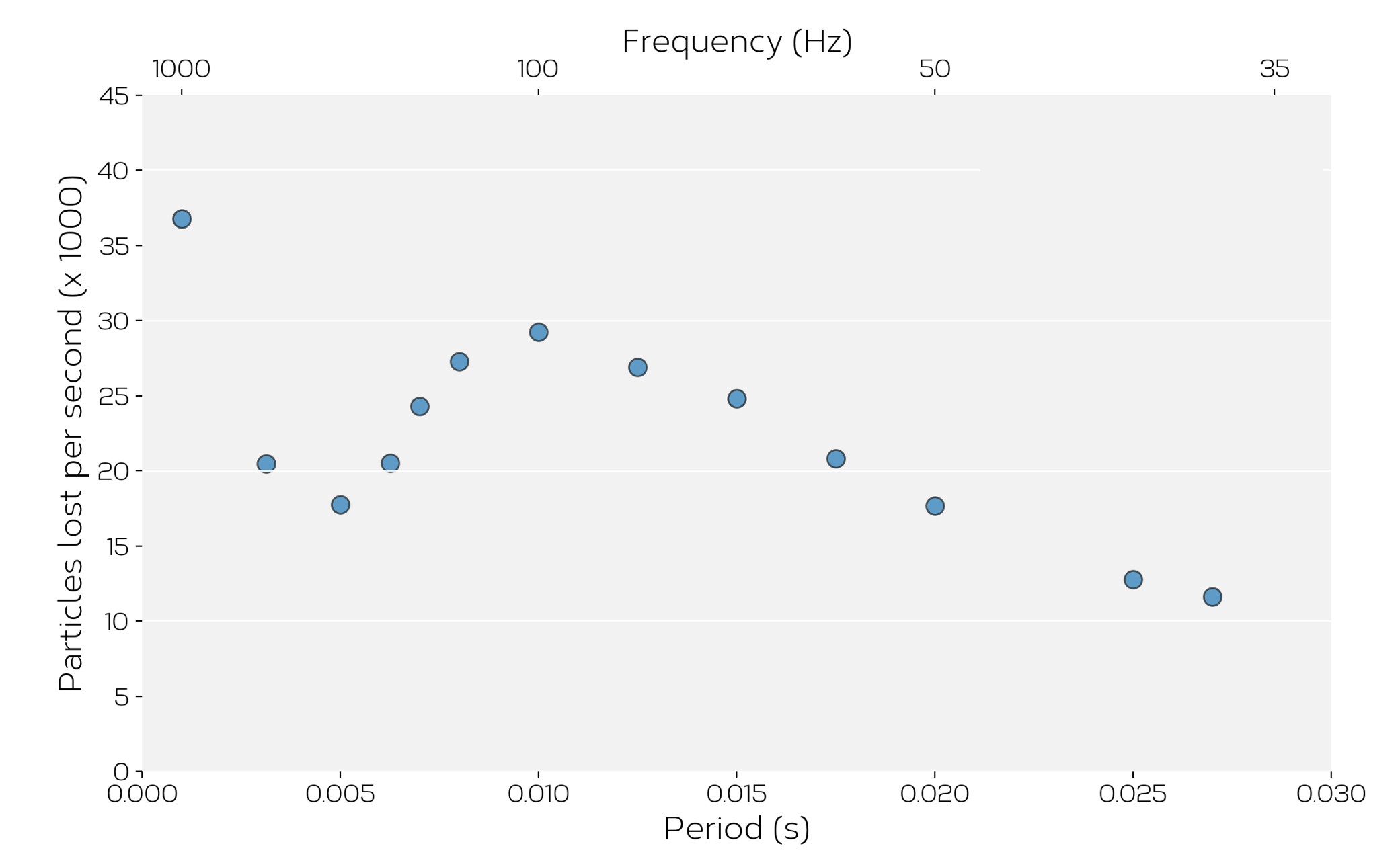}
\caption{The effect of feeder vibration frequency at fixed amplitude $A = 0.2$ mm on the feed rate of powder simulated via DEM. }
\label{fig:DEMfreq}
\end{figure}

While the experiments presented above were limited to displacement controlled conditions, one may wonder whether frequency control (at fixed displacement) is a viable option for controlling flow.  Figure \ref{fig:DEMfreq} shows that the relationship between feed rate and frequency is much more complex than between feed rate and amplitude (at fixed frequency) with the feed rate varying in a non-monotonic fashion with frequency.  This can be traced to effects similar to the one illustrated in figure \ref{fig:unstable} where, at characteristic frequencies depending on the displacement amplitude, the powder flow becomes unstable within the hopper.  This result suggests that controlling feeder flow at fixed frequency and taking advantage of the (nearly) linear relationship between flow and displacement amplitude should provide the best opportunity for process control.  It also points to the need for careful experimental calibration to identify conditions where unstable flow may be a concern.

A final comment on the potential impact of actual powders that are not monosized spheres but which, as discussed in reference to the two powders studied experimentally here, have size distributions, irregular shapes and/or a range of densities (e.g. if powder mixtures are used).  While the effect of particle shape has not been studied within the context of the DEM simulations presented here, the impact of a bi-modal distribution of particle sizes (at constant density) and particle density (at fixed size) has. Vibration induced segregation, often referred to as the `brazil nut effect', is known to lead to large particles being segregated to the top of a vertically vibrating granular bed\cite{Gray2018ParticleFlows}. A variety of views exist to explain this effect, but the controlling mechanisms remain a topic of active consideration \cite{Gray2018ParticleFlows, Jing2020RisingFlows}. Two additional sets of simulations were performed, following the same approach as described above, but with two significant differences.  In the first set of simulations, during feeding, particles with two distinct sizes (radii of 120~$\mu$m and 60~$\mu$m) were deposited into the feeder, all other particle properties being held constant.  In a second set of simulations, particles with two distinct densities (2.5 g/cm$^3$ or 9.0 g/cm$^3$ (two times smaller and larger than the densities of the particles used in the above simulations) were deposited all other particle properties, including particle size, being held the same as those used in previous simulations.  Also different from the above simulations was the fact that the powder was deposited onto a moving horizontal surface (properties the same as those used above) to simulate the deposition of a powder layer on a build table. 

\begin{figure}[htbp]
\centering
\includegraphics[width=\figwidth\textwidth]{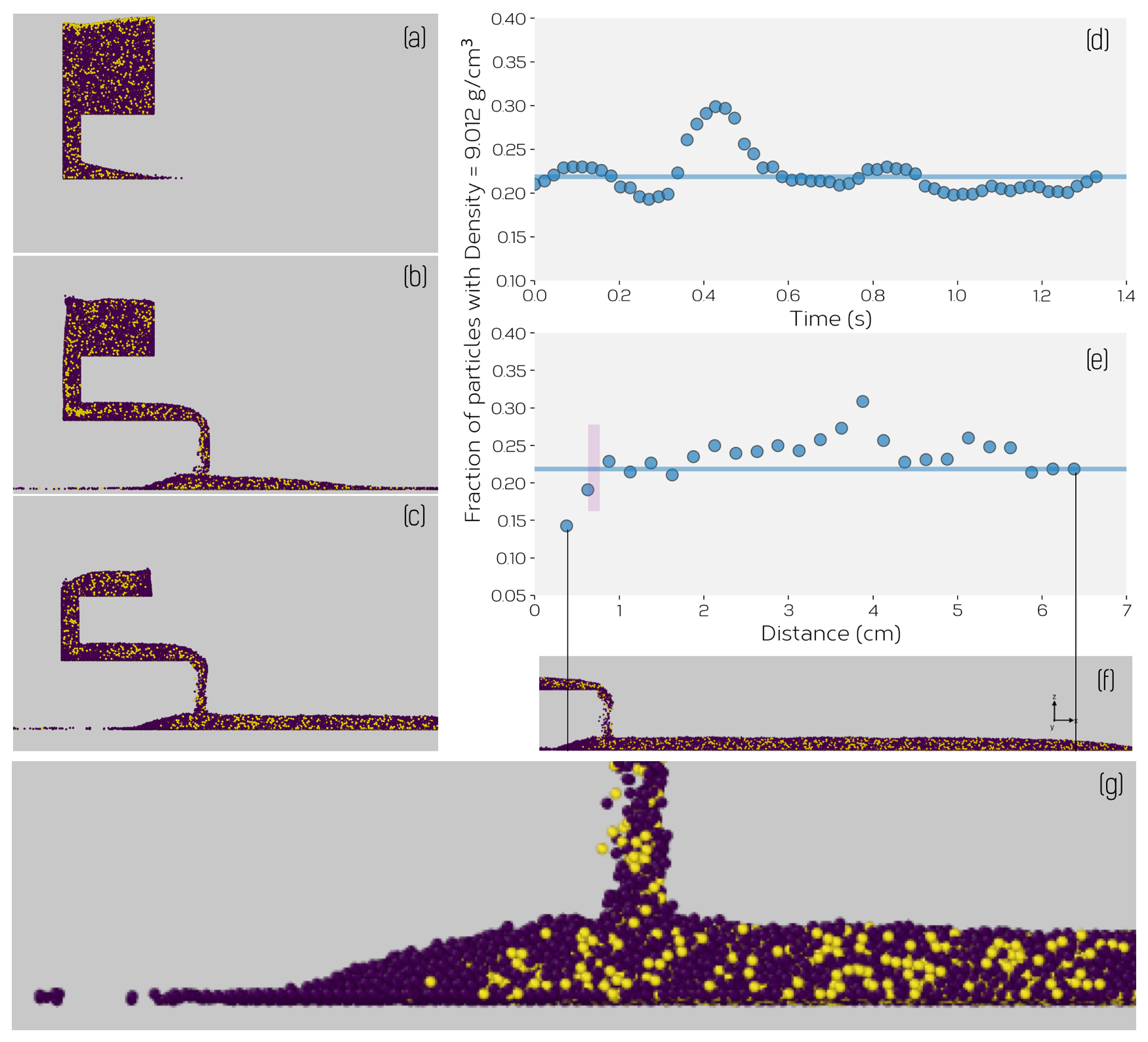}
\caption{Simulated powder bed feeding for a mix of two powders having the same size (radius of 120~$\mu$m) but two different densities. Panels a)-c) show the system before the onset of vibration, after 0.56~s and 1.1~s. Blue particles have a density of $\rho$ =2.25 g/cm$^3$ and yellow a density of 9.0 g/cm$^3$. d) Fraction of particles with $\rho$ =9.0 g/cm$^3$ exiting the hopper as a function of time ($t=0$s corresponding to the onset of feeding). e) The fraction of particles with a density of 9.0 g/cm$^3$ as a function of position along the powder layer, the powder layer position relative to this graph being shown in (f). The solid lines in (d) and (e) indicate the initial average fraction of particles with $\rho$ =9.0 g/cm$^3$. The vertical bar in (e) is intended to indicate the position of the feeder's exit chute. g) Shows a higher magnification of (f) illustrating the tendency for lower density particles to segregate to the right of the powder bed.}
\label{fig:twodensities}
\end{figure}

No evidence of density or size induced segregation was observed within the feeder for either of these simulations.  This can be illustrated by measuring the time evolution of the fraction of the particle types within the vertical channel separating the feed channel from the hopper as a function of time, using this as a measure of the `composition' of the powder being fed from the hopper.  This is illustrated for the case where two particle densities were used in figure \ref{fig:twodensities}d.  Note, the elevated fraction at $t \approx 0.4-0.5$ corresponds to the layer of higher density particles located at the top of the feeder (see e.g. figure \ref{fig:twodensities}a), this being an artifact arising from the `rain model' method used to fill the feeder.  The lack of discernible segregation based on size or density due to vibration is a consequence of the fact that any vibration induced convection within the hopper is overwhelmed by the downward flow of powder from the hopper into the feed channel.  Powder is drawn from across the entire width of the hopper into the feed channel with no evidence of preferential flow of one type of particle.

In the case of the powder bed itself, no evidence of segregation was observed for the case of the powder containing a bi-modal distribution of particle sizes. It was confirmed that the initial hopper composition and the composition of the powder bed (at all times) remained the same.  In the case of the powder with two particle densities, however, a small difference between the powder bed `composition' and bulk initial `composition' were observed, this being illustrated in figure \ref{fig:twodensities}e.  Careful inspection of figures \ref{fig:twodensities}e, f and g show that a preferential accumulation of lower density powder occurs to the left of the developing powder bed.  Due to the lower mass of these smaller particles, they tend to be ejected to a larger distance upon impact with the build table/particles on the build table after exiting the feeder.  This is much more obvious to the left of the feeder where the number of particles is lower and (over time) the particles occupying that space are themselves lower in mass. Ultimately, segregation reaches steady state leading to a stable powder bed composition slightly richer in the denser powder; 2\% more of the particles with  $\rho$ =9.0~g/cm$^3$ in the powder bed compared to than in the bulk.  The same 2\% enrichment of the powder bed was observed in a simulation where powder densities of $\rho$ =2.25 and 4.5 g/cm$^3$ were used.  

One way to further improve powder feeding, and to help further mix the powder as it enters the powder bed is to use the bottom surface of the feeder to `level' the powder bed.  In this mode the feeder is much closer to the build table and the powder `injected' into the powder bed, this leading to both a flatter surface and an additional opportunity for mixing of powders of different densities/sizes within the powder bed.  This opens up many further opportunities to think about alternative ways of delivering powder (or powders) during PBF-AM processing, with the potential to overcome many of the limitations discussed in the opening for conventional re-coating technologies.

\section{Conclusions}

In this work we have illustrated the use of a vibratory powder feeding system in conjunction with electron beam additive manufacturing.  It has been shown to be possible to feed not only `conventional' gas atomized powders used extensively in additive manufacturing but also a highly irregularly shaped and broadly sized water atomized Fe-Ni powder conventionally used in the powder metallurgy industry.  It has been shown that the feed rate, and thus the powder layer height can be accurately controlled thanks to a linear relationship between feed rate and vibration amplitude at fixed frequency.  Simulations have revealed the underlying mechanism controlling the feeding behaviour under these conditions including the threshold vibration amplitude below which no feeding occurs. Simulations suggest that size based segregation is not expected but that a small amount of powder based segregation may occur during feeding of particles with very different densities. Careful monitoring would be required in this case to ensure the desired bulk chemistry is obtained. Opportunities to introduce additional powder mixing in the powder bed itself have also been suggested.  

The use of a vibratory powder feeder, as introduced here, can overcome some of the limitations inherent in re-coaters, one significant one shown here being the ease with which it can distribute powder that would be considered to difficult to flow (without sizing or mixing with other powders) in conventional electron beam or laser based additive manufacturing.  This opens the opportunity for new strategies in powder delivery for additive manufacturing that are not conventionally considered in existing commercial systems.

\section{Bibliography}

\bibliography{powderfeedingDec32020-final}

\end{document}